# Performance Investigation of Virtual Private Networks with Different Bandwidth Allocations

Mahalakshmi Chidambara Natarajan[1], Ramaswamy Muthiah[2] and Alamelu Nachiappan[3]

[1,2] Electrical Engineering, Annamalai University
Annamalai Nagar – 608002, Tamil Nadu, India

[3] Electrical and Electronics Engineering, Pondicherry Engineering College
Puducherry – 605014, India

**Abstract**
A Virtual Private Network (VPN) provides private network connections over a publicly accessible shared network. The effective allocation of bandwidth for VPNs assumes significance in the present scenario due to varied traffic. Each VPN endpoint specifies bounds on the total amount of traffic that it is likely to send or receive at any time. The network provider tailors the VPN so that there is sufficient bandwidth for any traffic matrix that is consistent with these bounds. The approach incorporates the use of Ad-hoc On demand Distance Vector (AODV) protocol, with a view to accomplish an enhancement in the performance of the mobile networks. The NS2 based simulation results are evaluated in terms of its metrics for different bandwidth allocations, besides analyzing its performance in the event of exigencies such as link failures. The results highlight the suitability of the proposed strategy in the context of real time applications.

## 1. Introduction

Wireless utilities have become an essential part of communication networks. They are considered as an auxiliary approach to be used in regions where it is difficult to build a wired connection. The development is seen to progress at an astounding speed, with evidences of significant growth in the areas of mobile subscribers and terminals, mobile and wireless access networks, and mobile services and applications.

Wireless networks are primarily used in improvised environments where communication can be rapidly established between the nodes without requiring any fixed infrastructure. These networks emerge in a spontaneous manner when a node is within the transmission range of one or more nodes. The topology of the network is fundamentally fluid at all times. The routing protocols therefore act as the binding force to ensure connectivity between the nodes.

The inter-router protocols essentially have information about their neighboring routers. These routers perform the functions in a static manner to meet the needs of a link severing, balance the load and satisfy the quality of service requirements. The node neighborhood on the contrary, changes rapidly along with the link connectivity depending upon the wireless range and mobility of the network. Consequently, the routing information also needs to be precisely updated at a similar pace taking into account the present neighborhood information. Thus the accurate functioning of a wireless network is directly bound with the sincere and skilful execution of these routing protocols by the participating nodes.

The review revolves on the current work involved with Wireless Sensor Networks (WSN) as it is for the first time an attempt with VPN on similar lines. A distributed bandwidth guaranteed shortest path routing algorithm has been evolved in order to accomplish good performance [1]. A variable bandwidth allocation approach that uses time frequency slot assignment has been proposed to reduce energy consumption of a collaborative sensor network, with variation in node density at event rates [2]. A quantization algorithm that allows the best possible variance for a given bandwidth constraints has been presented [3]. A simple bandwidth management architecture that allows the network administrator to cater to the traffic generated in the sensor network has been developed [4]. A multicast routing protocol suitable for Wireless Sensor Network that ensures the tradeoff between optimality of the multicast tree and efficiency of the data delivery has been built [5]. A maximum throughput bandwidth allocation strategy suitable for multichannel wireless mesh network has been formulated to achieve a good tradeoff between fairness and throughput [6]. A load sharing scheme among sensor nodes that takes into account their computation capabilities and networking conditions has been suggested [7].

A Virtual Private Network (VPN) is a logical network that perceives data transmission through dedicated wireless means. The essential requirement for a VPN is to ensure the desired bandwidth, which demands the network provider to guarantee the VPN users that a certain amount of bandwidth will be available to them at any time, unaffected by the traffic sent through the physical network by other users (who are not part of the VPN). It is explicitly possible only if the stipulated bandwidth is reserved for the VPN on the links of the physical network. It is a significant task, from the perspective of the network provider to satisfy the bandwidth allocations for a VPN, while minimizing the amount of network resources that are assigned to it.

## 2. Problem Description

A deterministic data transfer Ad hoc on - demand Distance Vector (AODV) protocol, suitable for large VPNs is suggested in this paper. It is designed to reserve bandwidth in the network and route the traffic between endpoints in order that the reserved bandwidth supports every valid traffic matrix. It is proposed to build an algorithm that can result in significant capacity savings to the service provider, in such a way that the next better alternate path based on bandwidth allocation is chosen in the event of an exigency. The scheme is evaluated through performance indices that portray the efficiency of the topology.

## 3. Proposed Scheme

The scheme is designed so as to arrange a packet to traverse from the identified source to the desired destination. If a node does not have a valid route to a destination, it initiates a route discovery process by broadcasting a route request packet. This packet contains source and destination addresses along with a sequence number. The sequence numbers ensure that the routes are loop free and if the intermediate nodes reply to route requests, they reply with the latest information only. Each neighbor (if it is not aware of the destination) broadcasts this packet to its neighbors and this process is repeated till an intermediate node is reached that has recent route information or till it reaches the destination.

A model of the VPN as seen in figure 1 is considered in this paper for specifying the bandwidth requirements of a VPN. The idea is to specify for each VPN endpoint e, the maximum total bandwidth b+(e) of traffic that e will send into the network at any time and the maximum total bandwidth b−(e) of traffic that e will ever receive from the network at any time. The network capacity reserved for the VPN must be sufficient for every possible traffic pattern that is consistent with the b+ and b− values. The model provides a convenient way for the customers to specify their bandwidth requirements, but makes the problem of efficient bandwidth reservation harder than in the traditional model where the customer has to specify pairwise demands for all VPN endpoints.

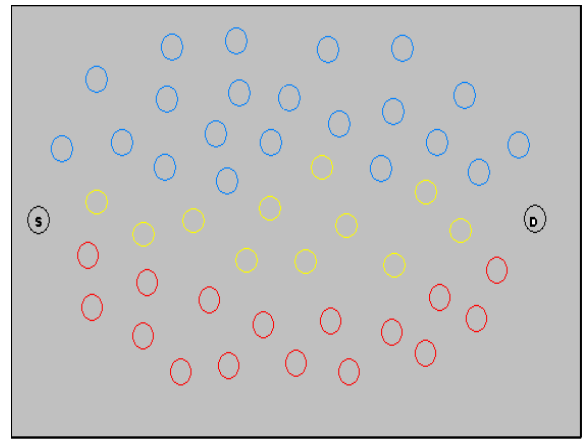

**Figure 1. Network Topology**

It is imperative that VPN traffic is effectively routed through the physical network, more so on a need based approach. It is in this direction that a specific scheme viz. Ad-hoc On-demand Distance Vector (AODV) is contemplated to accomplish the task with a perceived routing phenomenon. It is a routing protocol that is preferred particularly for dynamic link conditions. Every node in the network is designed to maintain a routing table, which contains information about the route to a particular destination. Whenever a packet is to be sent by a node, it first checks with its routing table to determine whether a route to the destination is already available. If so, it uses that route to send the packets to the destination. If a route is not available or the previously entered route is inactivated, then the node initiates a route discovery process. A Route REQuest packet (RREQ) is broadcasted by the node. Every node that receives the RREQ packet first checks if it is the destination for that packet and if so, it sends back a Route Reply packet (RREP). If it is not the destination, then it checks with its routing table to determine if it has got a route to the destination. If not, it relays the RREQ packet by broadcasting it to its neighbors. If its routing table does contain an entry to the destination, then the next step is the comparison of the 'Destination Sequence' number in its routing table to that present in the RREQ packet. This Destination Sequence number is the sequence number of the last sent packet from the destination to the source. If the destination sequence number present in the routing table is lesser than or equal to the one contained in the RREQ packet, then the node relays the request further to its neighbors. If the number in the routing table is higher than the number in the packet, it denotes that the route is a 'fresh route' and packets can be sent through this route. This intermediate node then sends a RREP packet to the node through which it received the RREQ packet. The RREP packet gets relayed back to the source through the reverse route. The source node then updates its routing table and sends its packet through this route. During the operation, if any node identifies a link failure it sends a Route ERRor packet (RERR) to all other nodes that uses this link for their communication to other nodes.

The minimum amount of bandwidth that must be reserved for the VPN on each of the links of the physical network is uniquely determined once the routing is specified. The challenge is to compute a good routing that minimizes the total amount of necessary bandwidth reservations. Therefore a protocol based approach suitable for multipath routings is adopted with a view to facilitate undeterred data transfer between the source and destination.

The bandwidth reserved for the VPN endpoint $e \in E$ is denoted by $x_e$. The philosophy is to find a valid reservation according to bandwidth allocation. The validity of a reservation depends on the routing. It is important to distinguish several routing paradigms that can be used to transport VPN traffic through the physical network. With multi-path routing, every pair $(u, v)$ is assigned a collection of paths from $u$ to $v$ together with a specification of which fraction of the traffic from $u$ to $v$ should be sent along each of these paths. For every pair $(u, v)$ of distinct nodes in $Q$, the traffic from $u$ to $v$ can be split arbitrarily among several paths. A bandwidth reservation $x$ is valid if and only if it holds for every valid traffic matrix $D$ and for every link $e \in E$ that

$$\sum_{u,v \in Q} d_{u,v} . f_{u,v} \leq x_e \qquad \text{(1)}$$

The term $\sum_{u,v \in Q} d_{u,v} . f_{u,v}$ in equation (1) expresses the amount of traffic that is routed through $e$ when the current traffic matrix is $D$.
However, with VPNs attempting to be wireless in their configuration it is the ardent desire to explore the performance with the role of AODV based strategy in the mission of transfer of information.

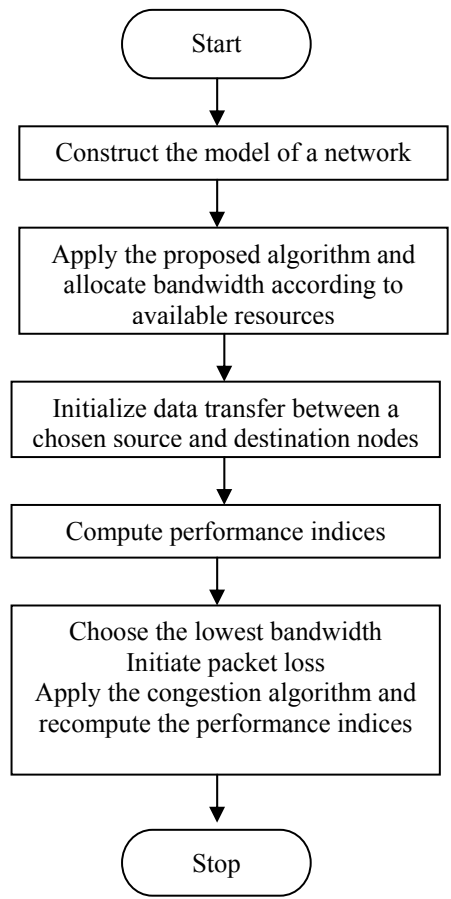

**Figure 2. Flow Chart for Bandwidth Allocation**

## 4. Simulation Results

The VPN is a 50-node random topology simulated using Network Simulator (NS2). The scheme envisages data transfer between a wireless source and destination. The first stage of the investigation is to allocate the baseline bandwidth restrictions in the three assumed routes as $2.1*10^6$, $1.8*10^6$, and $1.9*10^6$. The wireless nodes are configured to generate streams of traffic through a principle of fusing. A data message of size 512 bytes is to be transmitted. A comprehensive set of measurements of packets received, Packet Delivery Ratio (PDR), packet loss, routing delay and the energy expended through each of the available three paths are computed. Each reading is repeated a minimum of five times and the mean values of these samples are tabulated in Table 1.

It is observed that the path with minimum bandwidth enjoys the superior performance, in successfully transferring the most number of packets, with minimum loss and delay. It incurs minimum expenditure of energy besides offering the highest PDR. The other paths with the subsequent higher bandwidths accomplish their relative degrees of performance in accordance with the design of the algorithm.

**Table 1. Performance Indices**

| Path | Band width * 10⁶ | Packets Received | PDR | Routing Delay | Energy Consumed * 10³ | Packet Loss |
|---|---|---|---|---|---|---|
| 1 | 2.1 | 362 | 27 | 1.7 | 5.5 | 3.8 |
| **2** | **1.8** | **385** | **48** | **0.75** | **4.6** | **2.15** |
| 3 | 1.9 | 370 | 38 | 1.05 | 5 | 2.5 |

The results seen through figures 3 and 4 are evaluated through the same AODV protocol based approach and displayed through bar charts, when data are allowed to be transmitted between the same source and destination. The analysis serves to establish that the algorithm is consistent in its mission and reveals that the protocol is viable for large scale transmission.

The response of the network, with an increase in the number of packets transferred is investigated and seen from figure 3 that the minimum bandwidth path offers the highest increase in PDR with increase in the size of the packets contributes to enhance the energy efficiency and foster the cause of extending the lifetime of the network. The remaining two transitory paths do realize an increase in the PDR of relatively lower magnitudes.

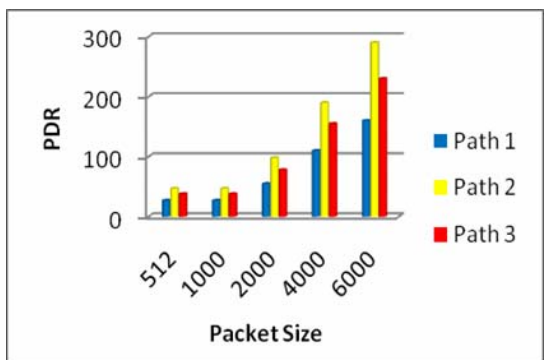

**Figure 3. Packet Delivery Ratio vs Packet Size**

The proposed scheme for an increase in the number of transmitted packets suffers from a loss of packets through transmission observed from figure 4. It is evident that a minimum bandwidth path retrogades the minimum loss while the other two hazzle with relatively higher losses.

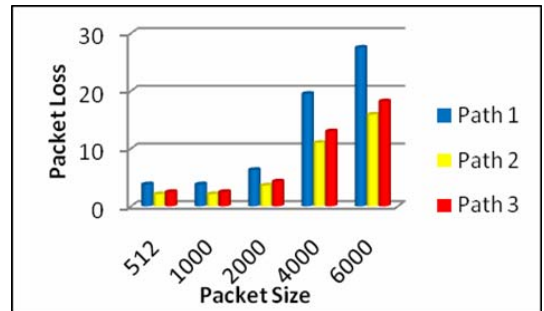

**Figure 4. Packet loss vs Packet Size**

The NS2 graphs obtained for the different performance indices in the path with the minimum bandwidth is displayed in figures 5 through 10. Figure 5 shows the usage of bandwidth with time in the chosen path.

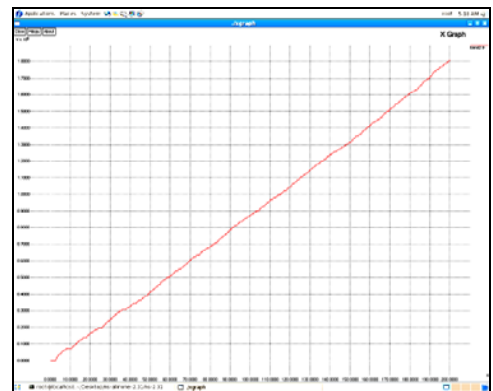

**Figure 5. Bandwidth vs Time**

The variation of number of packets received, PDR, energy consumed to accord data transmission, with respect to time are seen in figures 6, 7 and 8 respectively. The linear increase explains the suitability of the algorithm to handle varied traffic and cater to increases in accordance with the needs.

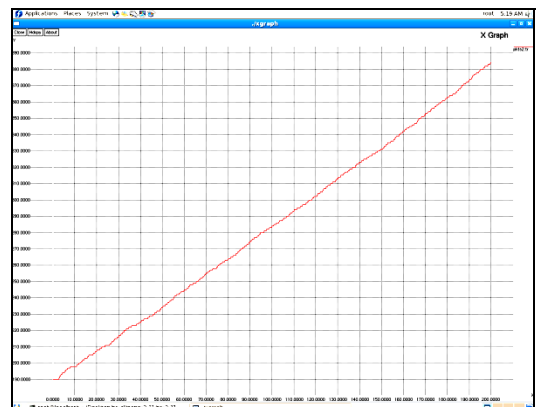

**Figure 6. Packets Received vs Time**

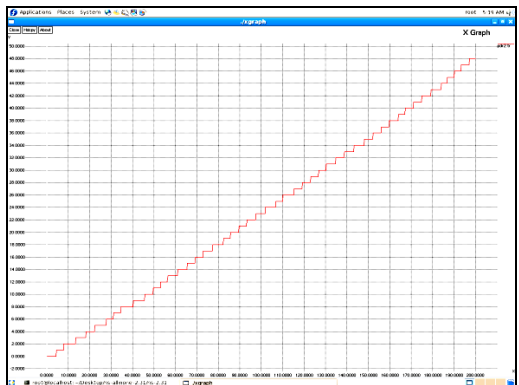

**Figure 7. Packet Delivery Ratio vs Time**

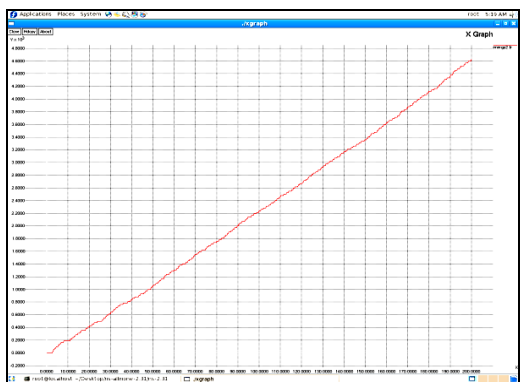

**Figure 8. Energy Consumed vs Time**

The routing delay of the nodes in the network when it encompasses data transmission is displayed in figure 9. It appears to be initially high for reasons of start up, but stabilizes thereafter and remains reasonably low throughout the linear increase in the packets transferred.

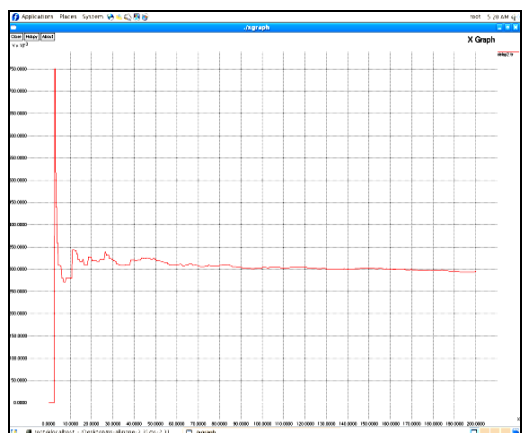

**Figure 9. Routing Delay vs Time**

The loss of packets increases with the increase in the number of packets transmitted as observed from figure 10. However, it is worthy to note that the loss is relatively low in this path when compared to other two paths as evident from Table 1.

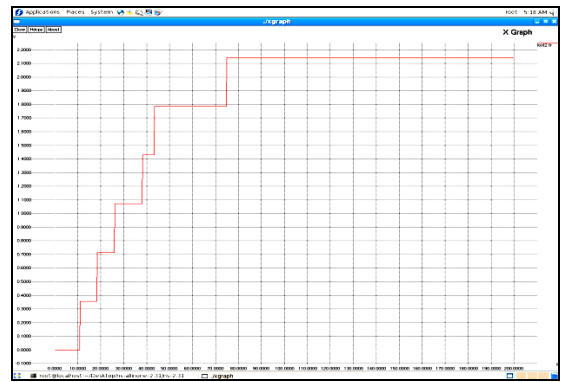

**Figure 10. Packet Loss vs Time**

In the event of the occurance of a link failure in the minimum bandwidth path, the network boils down to a two path topology. It is ensured that thereafter the performance is superior in the next minimumbandwidth path. The figures 11 and 12 relate the number of packets received and PDR to time to validate the design of the algorithm.

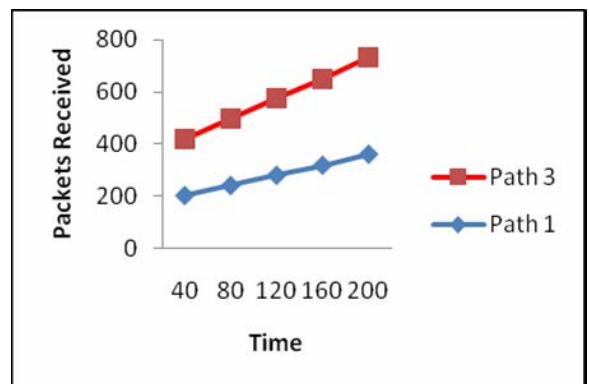

**Figure 11. Packets received for Alternate Paths**

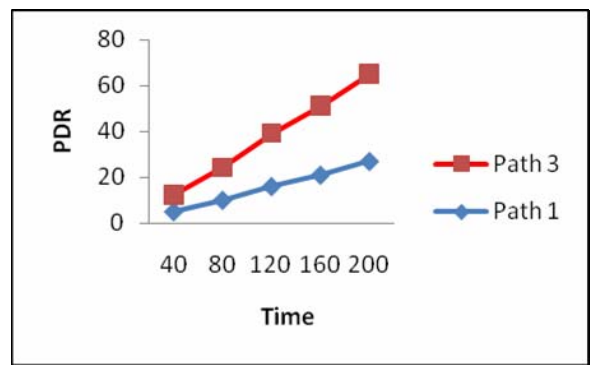

**Figure 12. PDR for Alternate Paths**

## 5. Conclusion

The AODV algorithm that can be applied to both symmetric and asymmetric bandwidth requirements has been designed to offer multipath routing and minimise the total cost of bandwidth reservation. It has been coined to handle natural constraints and inherit abilities to accordingly rearrange the topology to continue the process of data transmission. The feature that the network fuses in tune with the available resources has revealed its efficacy.

A bandwidth allocation scheme suitable for Virtual Private Network that enables wireless transmission has been developed. Its performance has been evaluated through simulation and the metrics computed. It has been constructed to extract the improved performance with minimum delay, packet loss and energy consumption. The results have brought out that the scheme has contributed to a reliable data transfer procedure that will considerably increase the life time of the network. The graphs have been projected to realize the efficiency of the new algorithm and its suitability for present day applications in real time networks. The fact that the proposed approach facilitates the best performance in the minimum bandwidth path will go a long way in ensuring optimal use of bandwidth in the present busy traffic scenario.

## Acknowledgment

The authors thank the authorities of Annamalai University for providing the necessary facilities in order to accomplish this piece of work.